\documentclass[10pt]{amsart}
\usepackage{amssymb}
\usepackage{enumerate}
\usepackage[dvips]{graphicx}
\DeclareGraphicsExtensions{.eps,.art,.ART,.ps}
\newcommand{\grad}{\operatorname{grad}}

\begin{document}
\begin{abstract}
We construct a coordinate system for the Kerr solution, based on the zero angular momentum observers dropped from infinity, which generalizes the Painlev\'{e}-Gullstrand coordinate system for the Schwarzschild solution. The Kerr metric can then be interpreted as describing space flowing on a (curved) Riemannian $3$-manifold. The stationary limit arises as the set of points on this manifold where the speed of the flow equals the speed of light, and the horizons as the set of points where the {\em radial} speed equals the speed of light. A deeper analysis of what is meant by the flow of space reveals that the acceleration of free-falling objects is generally {\em not} in the direction of this flow. Finally, we compare the new coordinate system with the closely related Doran coordinate system.
\end{abstract}
%
%
\title{Painlev\'{e}-Gullstrand Coordinates for the Kerr Solution}
\author{Jos\'{e} Nat\'{a}rio}
\address{Centro de An\'alise Matem\'atica, Geometria e Sistemas Din\^amicos, Departamento de Matem\'atica, Instituto Superior T\'ecnico, 1049-001 Lisboa, Portugal}
\thanks{This work was partially supported by the Funda\c{c}\~ao para a Ci\^encia e a Tecnologia
through the Program POCI 2010/FEDER and by grant POCI/MAT/58549/2004}
\maketitle
%
%
%
\section*{Introduction}
The Painlev\'e-Gullstrand coordinate system for the Schwarzschild solution \cite{Painleve21, Gullstrand22} is a particularly simple horizon-penetrating coordinate system, admitting interesting physical interpretations \cite{HL04, V04}. It was inexplicably overlooked for a long time, but its importance has increased in recent years due to its relation with analogue gravity models \cite{BLV05}.

In this paper we construct a generalization of the Painlev\'{e}-Gullstrand coordinate system for the Kerr solution \cite{Kerr63}, based on the zero angular momentum observers dropped from infinity. This is quite different from the generalization considered in \cite{HL04}, where the Doran form of the Kerr metric \cite{Doran00} was used.

The paper is organized as follows. In the first section we present the new coordinate system (but relegate the details of its derivation to an appendix so as not to interrupt the flow of the paper). In the second section we interpret this coordinate system as describing space flowing on a Riemannian $3$-manifod, with the stationary limit given by the set of points on this manifold where the speed of the flow equals the speed of light, and the horizons as the set of points where the {\em radial} speed equals the speed of light. In the third section we address the question of what is meant by the flow of space. In particular, we show that the acceleration of free-falling objects is generally {\em not} in the direction of this flow. We also show that motions close to the flow of space can be obtained from a classical conservative system with a magnetic term. Finally, in the fourth section we compare the new coordinate system with the Doran coordinate system.

We use the Einstein summation convention with latin indices $i,j,\ldots$ running from $1$ to $3$. Bold face letters ${\bf u}, {\bf v}, \ldots$ represent vectors on the space manifold.
%
%
%
\section{Coordinate system}
As we show in the appendix, the metric for the Kerr solution with mass $M>0$ and angular momentum $Ma$ can be written as
\[
ds^2 = -dt^2 + \frac{\rho^2}{\Sigma} \left( dr - v dt \right)^2 + \rho^2 d \theta^2 + \Sigma \sin^2 \theta \, (d \varphi + \delta d\theta - \Omega dt)^2,
\]
where the functions
\[
\rho^2 = r^2 + a^2 \cos^2 \theta \quad \text{ and } \quad \Sigma = r^2 + a^2 + \frac{2Mra^2}{\rho^2} \sin^2 \theta
\]
approach $r^2$ at infinity, the functions
\[
\Omega = \frac{2Mra}{\rho^2\Sigma} \quad \text{ and } \quad v = - \frac{\sqrt{2Mr(r^2 + a^2)}}{\rho^2}
\]
are the familiar expressions for the angular velocity and radial proper velocity of a zero angular momentum observer dropped from infinity \cite{Carter68}, and
\[
\delta = - a^2\sin(2\theta) \int_r^{+\infty} \frac{v \Omega}{\Sigma} dr.
\]
Here the coordinates $(t,r,\theta,\varphi)$ are related to the usual Boyer-Lindquist coordinates $(t',r,\theta,\varphi')$ by a coordinate transformation of the form
\[
\begin{cases}
t' = t + A(r) \\
\varphi' = \varphi + B(r,\theta)
\end{cases}
\]
where $A$ and $B$ are analytic functions with singularities at the horizons. The new coordinate $\varphi$ is introduced so that the new coordinate system penetrates the horizons (when they exist). Notice that for $a=0$ the coordinates $(t,r,\theta,\varphi)$ reduce to the usual Painlev\'{e}-Gullstrand coordinates for the Schwarzschild solution.
%
%
%
\section{Interpretation}
Following the usual interpretation of the Painlev\'{e}-Gullstrand coordinates \cite{HL04}, we can think of the Kerr solution as the Riemannian $3$-manifold with metric
\[
dl^2 = \frac{\rho^2}{\Sigma} dr^2 + \rho^2 d \theta^2 + \Sigma \sin^2 \theta \, (d \varphi + \delta d\theta)^2
\]
on which space is flowing with velocity 
\[
{\bf v} = v \frac{\partial}{\partial r} + \Omega \frac{\partial}{\partial \varphi}.
\]
As one would expect \cite{GP00, Kroon04}, the space manifold is not flat: for instance, the totally geodesic submanifold $\theta = \frac{\pi}2$ has Gauss curvature
\[
K = - \frac{3Ma^2}{r^5},
\]
which is strictly negative for $a \neq 0$. Notice that $K$ blows up at the singularity $\rho=0$.\footnote{It is interesting to note that near the singularity one has arbitrarily large circles of constant $(r, \theta)$, contained inside a region of finite volume (the volume element is simply $\rho^2 \sin \theta dr \wedge d\theta \wedge d\varphi$).}

The speed of the flow of space is given by the square root of
\[
\langle {\bf v}, {\bf v} \rangle = \frac{\rho^2}{\Sigma} \, v^2 + \Sigma \sin^2 \theta \, \Omega^2 = \frac{2Mr}{\rho^2}.
\]
Hence space is flowing at the speed of light on the stationary limit, determined by $1 - \frac{2Mr}{\rho^2}=0$. The {\em radial} speed, on the other hand, is given by the square root of
\[
\frac{\rho^2}{\Sigma} \, v^2 = 1 - \frac{\Delta}{\Sigma},
\]
where $\Delta = r^2 - 2Mr + a^2$, and equals the speed of light for $\Delta = 0$, i.e.~at the horizons (when they exist).

It is easily seen that
\[
ds^2 < 0 \Leftrightarrow \| {\bf u} - {\bf v} \| < 1,
\]
where
\[
{\bf u} = \frac{dr}{dt} \frac{\partial}{\partial r} + \frac{d\theta}{dt} \frac{\partial}{\partial \theta} + \frac{d\varphi}{dt} \frac{\partial}{\partial \varphi}
\]
is the velocity of a massive particle with respect to the time coordinate $t$. Therefore no observer inside the stationary limit can remain at rest. If $|a|<M$ then no observer inside the outer horizon can ever come out, and must indeed also cross the inner horizon; however, he does not necessarily have to wind up in the singularity $\rho=0$. The same is true for $|a|=M$ except that the two horizons now coincide.
%
%
%
\section{What is the flow of space?}
Stationary metrics of the form
\[
ds^2 = - dt^2 + \gamma_{ij} (dx^i - v^i dt) (dx^j - v^j dt)
\]
can be interpreted as suggesting that space (defined as the Riemannian $3$-manifold with metric\footnote{This metric corresponds to the local distances measured by the free-falling observers moving with the flow of space.} $dl^2=\gamma_{ij}dx^i dx^j$) is flowing with velocity ${\bf v} = v^i \frac{\partial}{\partial x^i}$. A simple but instructive example is Minkowski spacetime in an uniformly rotating frame with constant angular velocity $\omega$,
\[
ds^2 = - dt^2 + dr^2 + r^2(d \varphi + \omega dt)^2 + dz^2,
\]
for which the ``true'' (inertial) Euclidean $3$-space is flowing with velocity
\[
{\bf v} = - \omega \frac{\partial}{\partial \varphi}.
\]
This example shows that the acceleration of free-falling objects is generally {\em not} in the direction of the flow of space. In fact, it is possible to show that if we use the time coordinate $t$ as the parameter then the geodesic Lagrangian
\[
L = \sqrt{1 - \gamma_{ij} \left(u^i - v^i \right) \left(u^j - v^j \right)} \quad \quad \quad \left(u^i = \frac{dx^i}{dt}\right)
\]
leads to the equation of motion
\[
\frac{D}{dt} \left( \frac{u^i - v^i}{L} \right) + \nabla^i v_j  \left( \frac{u^j - v^j}{L} \right) = 0,
\]
where $\nabla$ is the Levi-Civita connection of the space manifold and $\frac{D}{dt}=u^i \nabla_i$ (compare with \cite{HL04}). If ${\bf u}={\bf 0}$, this becomes
\[
\frac{Du^i}{dt} + v^i v^j v^k \nabla_j v_k - v^j \nabla^i v_j = 0,
\]
which can be written as
\[
\frac{D {\bf u}}{dt} = - \grad \phi + \langle \grad \phi, {\bf v} \rangle {\bf v},
\]
where $\phi = - \frac12 \| {\bf v} \|^2$. Therefore the acceleration of a particle at rest is primarily in the direction along which $ \| {\bf v} \|^2$ grows maximally, not in the direction of ${\bf v}$ (the second term in the acceleration only becomes important for $\|{\bf v}\| \sim 1$, i.e.~near the stationary limit). For Minkowski spacetime in a rotating frame, for instance, the centrifugal acceleration felt by a stationary particle is in the radial direction.

Notice by the way that motions which a close to the flow of space (in the sense that $\|{\bf u}-{\bf v}\| \ll 1$) are given by the approximate Lagrangian
\[
L = \frac12 \gamma_{ij} \left(u^i - v^i \right) \left(u^j - v^j \right) = \frac12 \|{\bf u}\|^2 + \mu({\bf u}) - \phi,
\]
where $\mu({\bf u}) = - \langle {\bf v}, {\bf u} \rangle$. This Lagrangian defines a conservative system with magnetic term on the space manifold. If in particular $\mu$ is closed we obtain a conservative system, which can be interpreted as a Newtonian gravitational field \cite{N06}. This is not the case for the Kerr solution, for which
\[
\mu = \frac{\rho^2}{\Sigma} \, v \, dr + \Sigma \sin^2 \theta \, \Omega \, (d \varphi + \delta d\theta)
\]
is readily shown not to be closed. In other words, the flow of space is not irrotational in this case.\footnote{In this respect it is misleading to compare the flow of space in the Kerr solution to the flow of water draining from a bathtub. This may seem strange in light of the fact that the zero angular momentum observers dropped from infinity have zero vorticity, but one must bear in mind that stationary observers have themselves nonzero vorticity.}
%
%
%
\section{Relation to the Doran coordinates}
This new coordinate system is, of course, closely related to the Doran coordinate system $(t,r,\theta,\varphi'')$, in terms of which the Kerr metric is written \cite{Doran00}
\[
ds^2 = - dt^2 + \frac{\rho^2}{r^2 + a^2} \left[ dr - v (dt - a \sin^2 \theta d\varphi'')\right]^2 + \rho^2 d\theta^2 + (r^2 + a^2) \sin^2 \theta {d\varphi''}^2.
\]
As suggested by the notation, only the azimuthal angular coordinate differs in the two coordinate systems. They are related to the Boyer-Lindquist azimuthal angular coordinate $\varphi'$ by
\[
d \varphi' = d \varphi + \alpha \Omega dr + \delta d \theta = d \varphi'' + \frac{\alpha a}{r^2 + a^2} dr
\]
where the function $\alpha=\alpha(r)$ is defined in the appendix. The coordinate $\varphi''$ is simply the initial Boyer-Lindquist coordinate $\varphi'$ of the zero angular momentum observers dropped from infinity, which leads to the interpretation that in the Doran coordinate system these observers fall radially \cite{HL04}.

Although the Doran coordinate system appears simpler, its interpretation is not as straightforward, since one must consider a ``twist field'' in addition to the velocity field \cite{HL04}. This extra piece of information is coded in the curvature of the space manifold in our new coordinate system.
%
%
%
\section*{Appendix}
Here we show how to write the Kerr metric in the Painlev\'{e}-Gullstrand form. In the usual Boyer-Lindquist coordinates $(t',r,\theta,\varphi')$, the metric is written \cite{ONeill95}
\[
ds^2 = - \left( 1 - \frac{2Mr}{\rho^2} \right) d{t'}^2 + \frac{\rho^2}{\Delta} dr^2 + \rho^2 d \theta^2 + \Sigma \sin^2 \theta d{\varphi'}^2 - \frac{4Mra\sin^2 \theta}{\rho^2} dt' d\varphi'.
\]
To obtain the Painlev\'{e}-Gullstrand coordinates $(t,r,\theta,\varphi)$ we perform the coordinate transformation defined by
\begin{equation} \label{coordinates}
\begin{cases}
dt' = dt + \alpha(r, \theta) dr + \gamma(r, \theta) d\theta \\
d\varphi' = d\varphi + \beta(r,\theta) dr + \delta(r, \theta) d\theta 
\end{cases}
\end{equation}
(where $\alpha, \beta, \gamma$ and $\delta$ are functions to be determined), leading to
\begin{align} \label{metric}
ds^2 = & - \left( 1 - \frac{2Mr}{\rho^2} \right) (dt + \alpha dr + \gamma d\theta)^2 + \frac{\rho^2}{\Delta} dr^2 \\
& + \rho^2 d \theta^2 + \Sigma \sin^2 \theta (d\varphi + \beta dr + \delta d\theta)^2 \nonumber \\
& - \frac{4Mra\sin^2 \theta}{\rho^2} (dt + \alpha dr + \gamma d\theta)(d\varphi + \beta dr + \delta d\theta). \nonumber
\end{align}
Requiring the coefficient of $dr d\varphi$ to vanish yields
\begin{equation} \label{beta}
\beta = \Omega \alpha.
\end{equation}
This implies that the angular velocity for a zero angular momentum observer dropped from infinity in the new coordinates is still given by $\Omega$. Assuming that the radial proper velocity in the new coordinates is also still given by $v$, we know that the metric in the new coordinates must be of the form
\begin{align} \label{Painleve}
ds^2 = & - dt^2 + \gamma_{rr} (dr - vdt)^2 + 2 \gamma_{r \theta} (dr - vdt) d\theta \\
& + \gamma_{\theta\theta} d\theta^2 + \Sigma \sin^2 \theta (d\varphi - \Omega dt)^2 + 2 \gamma_{\theta \varphi} d\theta(d\varphi - \Omega dt). \nonumber
\end{align}
Equating the coefficients of $dt^2$ in \eqref{metric} and \eqref{Painleve} yields
\[
\gamma_{rr}=\frac{\rho^2}{\Sigma}.
\]
From the equality of the coefficients of $dtdr$ we obtain
\begin{equation} \label{alpha}
\alpha = - \frac{\sqrt{2Mr(r^2 + a^2)}}{\Delta}.
\end{equation}
In particular, $\alpha$ depends only on $r$. Since $\alpha dr + \gamma d \theta$ must be exact, we see that $\gamma$ can depend only on $\theta$; if we take $\gamma$ to vanish for $r=+\infty$ then we have $\gamma=0$.\footnote{This fixes the time coordinate to coincide with the Doran time coordinate.} Notice that \eqref{alpha} also determines $\beta$ through \eqref{beta}. Since $\beta dr + \delta d \theta$ must be exact, we obtain
\[
\delta = - \int_r^{+\infty} \frac{\partial \beta}{\partial \theta} dr = - \int_r^{+\infty} \alpha \frac{\partial \Omega}{\partial \theta} dr = - \int_r^{+\infty} \frac{\rho^2v}{\Delta} \frac{\partial \Omega}{\partial \theta} dr = - a^2\sin(2\theta) \int_r^{+\infty} \frac{v \Omega}{\Sigma} dr,
\]
(where again we choose $\delta$ to vanish for $r=+\infty$). The condition that the coefficients of $dr^2$ are the same yields
\[
\alpha^2 = \frac{2Mr(r^2 + a^2)}{\Delta^2},
\]
in agreement with \eqref{alpha}. Equating the coefficients of $dr d\theta$ leads to
\[
\gamma_{r \theta} = 0.
\]
The coefficients of $dtd\varphi$ are automatically equal, and for the coefficients of $dtd\theta$ to coincide we must have
\begin{equation} \label{one}
\gamma_{\theta\varphi} = \Sigma \sin^2 \theta \, \delta,
\end{equation}
which is exactly the condition for the coefficients of $d \theta d \varphi$ to be equal. Finally, the equality of the coefficients of $d\theta^2$ implies
\[
\gamma_{\theta\theta} = \rho^2 + \Sigma\sin^2\theta \, \delta^2. 
\]
This completes the matching of the coefficients of \eqref{metric} and \eqref{Painleve}. Since we did not reach any contradictions (which was by no means trivial, since several conditions from the matching of the coefficients turned out to be automatically satisfied), the coordinate transformation \eqref{coordinates} does lead to the Painlev\'{e}-Gullstrand form of the metric:
\[
ds^2 = -dt^2 + \frac{\rho^2}{\Sigma} \left( dr - v dt \right)^2 + \rho^2 d \theta^2 + \Sigma \sin^2 \theta \, (d \varphi + \delta d\theta - \Omega dt)^2.
\]
Notice that although $\alpha$ and $\beta$ blow up when $\Delta$ vanishes, $\delta$ does not, and hence this coordinate system penetrates the horizons (when they exist).
%
%
%

\end{document}